\documentclass[%
reprint,
amsmath,amssymb,
aps,
]{revtex4-1}

\usepackage{graphicx}
\usepackage{dcolumn}
\usepackage{bm}
\usepackage{enumerate}
\usepackage{bbm}
\usepackage{float}
\usepackage{tabularx,booktabs}
\newcolumntype{Y}{>{\centering\arraybackslash}X}
\usepackage[subpreambles=true]{standalone}
\usepackage{import}
\usepackage[ruled,vlined]{algorithm2e}
\newtheorem{proposition}{Proposition}
\newtheorem{assumption}{Assumption}
\newtheorem{lemma}{Lemma}
\usepackage[mathlines]{lineno}

\begin{document}


\title{The Wang-Landau Algorithm as Stochastic Optimization and Its Acceleration}

\author{Chenguang Dai}
\email{chenguangdai@g.harvard.edu}
\author{Jun S. Liu}
\email{jliu@stat.harvard.edu}
\affiliation{Department of Statistics, Harvard University}%

\date{\today}

\begin{abstract}
We show that the Wang-Landau algorithm can be formulated as a stochastic gradient descent algorithm minimizing a smooth and convex objective function, of which the gradient is estimated using Markov chain Monte Carlo iterations. The optimization formulation provides us a new way to establish the convergence rate of the Wang-Landau algorithm, by exploiting the fact that almost surely, the density estimates (on the logarithmic scale) remain in a compact set, upon which the objective function is strongly convex.
The optimization viewpoint motivates us to improve the efficiency of the  Wang-Landau algorithm using popular tools including the momentum method and the adaptive learning rate method. We demonstrate the accelerated Wang-Landau algorithm on a two-dimensional Ising model and a two-dimensional ten-state Potts model. 
\end{abstract}

\pacs{Valid PACS appear here}
\maketitle

\section{Introduction}
\label{sec:intro}
The Wang-Landau (WL) algorithm \citep{wang2001efficient, wang2001determining, landau2004new} has been proven useful in solving a wide range of computational problems in statistical physics, including spin-glass models \citep{brown2005wang, torbruegge2007sampling, alder2004dynamics, snider2005absence, okabe2002application, zhou2006wang, wu2005ground, malakis2006lack, hernandez2008wang, fytas2008phase, tsai2007critical, yamaguchi2001three}, fluid phase equilibria \citep{mastny2005direct, shell2002generalization}, polymers \citep{taylor2009phase, strathmann2008transitions}, lattice gauge theory \citep{langfeld2012density}, protein folding \citep{rathore2002monte, rathore2003density, rathore2004molecular}, free energy profile \citep{calvo2002sampling}, and numerical integration \citep{troster2005wang, li2007numerical}. Its successful applications in statistics have also been documented \citep{liang2005generalized,atchade2010wang,bornn2013adaptive}. The WL algorithm directly targets the density of states (the number of all possible configurations for an energy level of a system), thus allowing us to calculate thermodynamic quantities over an arbitrary range of temperature within a single run of the algorithm.

Much effort has been made to understand the dynamics of the WL algorithm, along with numerous proposed improvements, of which we highlight three here. 
(i) Optimizing the modification factor (flatness criterion) \citep{belardinelli2007fast, zhou2005understanding, zhou2008optimal, dayal2004performance}. Belardinelli and Pereyra \citep{belardinelli2007fast} proposed that instead of reducing the modification factor exponentially, the log modification factor should be scaled down at the rate of $1/t$ in order to avoid the saturation in the error. (ii) Employing a Parallelization scheme. Wang and Landau \citep{wang2001efficient} suggested that multiple random walkers working simultaneously on the same density of states can accelerate the convergence of the WL algorithm. The efficiency of the parallelization scheme can be further enhanced using the replica-exchange framework \citep{vogel2013generic}. 
(iii) Incorporating efficient Monte Carlo trial moves \citep{wust2009versatile, yamaguchi2002combination, wu2005overcoming}. 

In this paper, we consider the WL algorithm from an optimization perspective and formulate it as a first-order method. We derive the corresponding smooth and convex objective function, of which the gradient involves the unknown density of states. 
Wang and Landau \citep{wang2001efficient} used a random-walk based Metropolis algorithm \cite{metropolis1953equation}
to estimate the gradient. In general, any suitable Markov chain Monte Carlo (MCMC) strategies \cite{liu2008monte} can be employed for this purpose. Therefore, the WL algorithm is essentially a stochastic gradient descent algorithm. 

The optimization viewpoint enables us to establish the convergence rate of the WL algorithm. Following \citep{fort2015convergence} and using the standard stochastic approximation theory \citep{fort2011convergence}, we first show that the density estimates (on the logarithmic scale) almost surely stay in a compact set. Based on this, we exploit the strong convexity of the objective function, restricted on this compact set, to prove the convergence rate. We note that the gradient estimator output from the MCMC iterations is generally biased, thus a critical step is to show that the bias vanishes properly as $t\to\infty$.

The optimization framework also provides us with a new direction for improving the WL algorithm. We explore one possible improvement, by combining the momentum method \citep{polyak1964some} and the adaptive learning rate method \citep{duchi2011adaptive, zeiler2012adadelta}. The general goal is to accelerate the transient phase \citep{darken1992towards} of the WL algorithm before it enters the fine local convergence regime. The effectiveness of the acceleration method is demonstrated on a two-dimensional Ising model and a two-dimensional ten-state Potts model, in which the learning in the transient phase is considerably demanding.

The rest of the paper is organized as follows. Section \ref{sec:WL} discusses the optimization formulation of the WL algorithm, and establishes the convergence rate from an optimization perspective.  Section \ref{sec:acceleration}  introduces possible strategies to accelerate the WL algorithm using optimization tools. Section \ref{sec:illustration} demonstrates the accelerated WL algorithm on two benchmark examples. Finally, Section \ref{sec:conclusion}  concludes with a few remarks. 

\section{An Optimization Formulation}
\label{sec:WL}
Let the space of all microscopic configurations be $\mathsf{X}$. Suppose there are totally $N$ energy levels, $E_1 < \cdots < E_N$, for the underlying physical model. For a microscopic configuration $x\in\mathsf{X}$, we use $E(x)$ to denote its energy. Let $\{g(E_n)\}_{n = 1}^N$ be the normalized density of states, i.e.,
\begin{equation}
g(E_n) \propto \#\{x\in\mathsf{X}, E(x) = E_n\},\ \ \ \sum_{n = 1}^N g(E_n) = 1.
\end{equation}
After initializing $g_0(E_n)$ as $1/N$, the WL algorithm iterates between the following two steps: (i) Propose a transition configuration and accept it with probability $\min\{1, g_t(E_i)/g_t(E_j)\}$, where $E_i$ and $E_j$ refer to the energy levels before and after this transition, respectively. This is essentially a step of the Metropolis algorithm \citep{metropolis1953equation} with the corresponding stationary distribution: 
\begin{equation}
\label{eq:intermediate-target}
\pi_t(x) \propto \sum_{n = 1}^N \frac{1}{g_t(E_n)}\mathbbm{1}\left(E(x) = E_n\right).
\end{equation}
(ii) Update the density of states. If $E(x_{t + 1}) = E_n$, multiply $g_t(E_n)$ by a modification factor $f_{t + 1} > 1$. That is, $g_{t + 1}(E_n) \leftarrow g_t(E_n) \times f_{t + 1}$. 

The modification factor $f_t$ should be properly scaled down in order to guarantee the convergence of the algorithm. There is a rich literature on how to adapt  $f_t$ online, including the flat/minimum histogram criterion,
and the $1/t$ rule \citep{belardinelli2007fast} with its various extensions \citep{jayasri2005wang, poulain2006performances}. Under a proper scaling rule, the magnitude of the modification factor $f_t$ is informative of the estimation error \citep{zhou2005understanding}. Thus, a commonly used stopping criteria for the WL algorithm is that $f_t$ is small enough (say, below $\exp(10^{-8})$).  

In the following, we will work on the logarithmic scale of the density of states. Denote $u_n^{(t)} = \log(g_t(E_n))$ for $n \in [N]$, and let $\bm{u} = (u_1,\cdots, u_N)$. 
The density update in the WL algorithm can be rewritten as
\begin{equation}
\label{eq:density-update}
u_n^{(t + 1)} \leftarrow u_n^{(t)} + \eta_{t + 1}\mathbbm{1}(E(x_{t + 1}) = E_n),
\end{equation}
where $\eta_{t + 1} = \log f_{t + 1}$, which will be referred to as the learning rate henceforth. The intermediate target distribution $\pi_t(x)$ defined in Equation \eqref{eq:intermediate-target} can also be formulated in terms of $\bm{u}^{(t)}$. We define
\begin{equation}
\label{eq:log-intermediate-target}
\pi_{\bm{u}}(x) \propto \sum_{n = 1}^N\exp(-u_n)\mathbbm{1}\left(E(x) = E_n\right),
\end{equation}
and denote $P_{\bm{u}}$ as a general transition kernel invariant to $\pi_{\bm{u}}(x)$.
For notational convenience, we use $\pi_t(x)$ to refer to  $\pi_{\bm{u}^{(t)}}(x)$, and use $P_t$ to refer to the transition kernel invariant to $\pi_t(x)$. After each density update, we normalize $\bm{u}^{(t)}$ to sum to 0, i.e.,
$u_n^{(t)} \leftarrow u_n^{(t)} - \sum_{i = 1}^N u_i^{(t)}/N$,
so that $\bm{u}^{(t)}$ stays in a compact set (see Proposition \ref{prop:convergence-WL}).
The WL algorithm can be slightly rephrased as in Algorithm~\ref{alg:WL}.
\begin{algorithm}
\label{alg:WL}
\caption{The Wang-Landau algorithm}
\begin{enumerate}
\item Initialization. $u^{(0)}_n = 0$ for $n\in[N]$.
\item For $t \geq 1$, iterate between the following steps.
\begin{enumerate}
\item Sample $x_{t + 1}$ from $P_t(x_t, \cdot)$.
\item Update $\bm{u}^{(t + 1)}$ following Equation \eqref{eq:density-update}.
\item Normalize $\bm{u}^{(t + 1)}$ to sum to 0.
\item Scale down the learning rate $\eta_t$ properly.
\end{enumerate}
\item Stop when the learning rate $\eta_t$ is smaller than a prescribed threshold.
\end{enumerate}
\end{algorithm}

Let us consider the following optimization problem:
\begin{equation}
\label{eq:optimization}
\begin{aligned}
& \min_{\bm{u} \in \mathbbm{R}^N} h(\bm{u}) = \log\left(\sum_{n = 1}^N  \exp(u^\star_n - u_n)\right),\\
& \text{subject to}\ \ \sum_{n = 1}^N u_n = 0,
\end{aligned}
\end{equation}
in which
$
u_n^\star = \log(g(E_n)) - \frac{1}{N}\sum_{i = 1}^N\log(g(E_i))$.
We write $\bm{u}^\star = (u^\star_1, \cdots, u^\star_N)$.
It is not difficult to see that this is a convex optimization problem because the objective function $h(\bm{u})$ is a log-sum-exp function and the constraint is linear.
It has a unique solution at $u_n = u^\star_n$ for $n\in[N]$, in which $\exp(u^\star_n)$ equals to the density of states $g(E_n)$ up to an multiplicative constant.

The projected gradient descent algorithm is a standard approach to solve the constrained optimization problem \eqref{eq:optimization}. The gradient of the objective function $h(\bm{u})$ is
\begin{equation}
\label{eq:gradient}
\frac{\partial h(\bm{u})}{\partial u_n} 
= -\frac{\exp\left(u_n^\star - u_n\right)}{\sum_{i = 1}^N \exp\left(u_i^\star - u_i\right)},\ \ \ n\in[N],
\end{equation}
which is not directly available because it involves the unknown density of states. However, one can think of approximating the gradient function defined in Equation \eqref{eq:gradient} by one-step or multiple-step Monte Carlo simulations, leading to a stochastic version of the projected gradient descent algorithm.

More precisely, a gradient descent step for minimizing $h(\bm{u})$ takes the following form:
\begin{equation}
\label{eq:update2}
u_n^{(t + 1)} \leftarrow u_n^{(t)} +  \frac{\eta_{t + 1}\exp(u^\star_n - u_n^{(t)})}{\sum_{i = 1}^N \exp(u^\star_i - u_i^{(t)})}.
\end{equation}
Denote the probability of the set $\{x\in\mathsf{X}: E(x) = E_n\}$ with respect to $\pi_t(x)$ as $\pi_t(E_n)$. Since the probability $\pi_t(E_n)$ is proportional to $\exp(u^\star_n - u_n^{(t)})$, the density update in Equation \eqref{eq:update2} is essentially
\begin{equation}
u^{(t + 1)}_n \leftarrow u_n^{(t)} + \eta_{t + 1}\pi_t(E_n).
\label{eq:update1}
\end{equation}
A crude approximation to $\pi_t(E_n)$ is the indicator function $\mathbbm{1}\left(E(x_{t + 1}) = E_i\right)$, given that after several steps of Monte Carlo simulations according to the transition kernel $P_t$ invariant to $\pi_t(x)$, $x_{t + 1}$ is approximately a sample from $\pi_t(x)$. This corresponds to the density update in Equation \eqref{eq:density-update}.

We note that the projection step to the set $\Pi = \{\bm{u}\in\mathbbm{R}^N, \sum_{n = 1}^N u_n = 0\}$ is equivalent to the normalization step (see Algorithm \ref{alg:WL} step 2(c)).
Thus, we have shown that the stochastic projected gradient descent algorithm solving the constrained optimization problem \eqref{eq:optimization}, which estimates the probability $\pi_t(E_n)$ by $\mathbbm{1}\left(E(x_{t + 1}) = E_n\right)$ using the output from Monte Carlo simulations, is equivalent to the WL algorithm.

The above optimization formulation has the following immediate implications. 
First, the parallel WL algorithm estimates the negative gradient $\pi_t(E_n)$ by $1/m\sum_{k = 1}^m[\mathbbm{1}(E(x^{(k)}_t) = E_n)]$, in which $m$ denotes the total number of random walkers, and $x^{(k)}_t$ denotes the $k$th random walker. Therefore, it reduces the variance of the gradient estimate by a factor $m$. Second, instead of implementing a single transition step, the separation strategy mentioned in \citep{zhou2005understanding} implements multiple transition steps within each iteration, so that the law of the random walker gets closer to the intermediate target distribution $\pi_t(x)$ defined in Equation \eqref{eq:log-intermediate-target}. Therefore, it reduces the bias of the gradient estimate. 

The optimization formulation also points out a new approach to establish the convergence rate of the WL algorithm. We first state a required assumption, which assumes that the transition kernels are (uniformly) geometrically ergodic over the space $\Pi$.
\begin{assumption}
\label{assump:uniform-ergodicity}
There exists a constant $\rho \in (0, 1)$ such that for all $\bm{u}\in\Pi$, $x\in\mathsf{X}$, $k\in\mathbbm{N}$, we have
\begin{equation}
\sup_{\bm{u}\in \Pi}\sup_{x\in\mathsf{X}}||P^k_{\bm{u}}(x, \cdot) - \pi_{\bm{u}}||_{\textnormal{TV}} \leq 2(1 - \rho)^k,
\end{equation}
in which for a signed measure $\mu$, the total variation norm is defined as
\begin{equation}
||\mu||_{\textnormal{TV}} = \sup_{|q|\leq 1}\left|\int_{\mathsf{X}}q(x)\mu(dx)\right|.
\end{equation}
\end{assumption}
We note that sufficient conditions for Assumption \ref{assump:uniform-ergodicity} exist in the literature (e.g., condition A2 in \citep{fort2015convergence}), and relaxation of Assumption \ref{assump:uniform-ergodicity} is also possible \citep{fort2011convergence}. 
We have the following result.
\begin{proposition}
\label{prop:convergence-WL}
Under Assumption \ref{assump:uniform-ergodicity}, if we scale down the learning rate $\eta_t$ in the order of $O(1/t)$, the following two statements hold.
\begin{enumerate}
\item Almost surely convergence.
\begin{enumerate}
\item There exists a compact set $\mathcal{K} \subseteq \Pi$ such that for any $t \geq 0$, $\bm{u}^{(t)} \in \mathcal{K}$ almost surely.
\item $\mathbbm{P}(\lim_{t\to\infty} \bm{u}^{(t)} = \bm{u}^\star) = 1$.
\end{enumerate}
\item Convergence rate.
There exists a constant $C > 0$ such that
\begin{equation}
\mathbbm{E}||\bm{u}^{(t)} - \bm{u}^\star||^2 \leq C/t.
\end{equation}
\end{enumerate}
\end{proposition}
The proof of Proposition \ref{prop:convergence-WL} is given in the Supplemental Material.

The first part of Proposition \ref{prop:convergence-WL} follows similarly as \citep{fort2015convergence}. The main idea is to rewrite the WL update, including the density update and the normalization step, as 
\begin{equation}\nonumber
\bm{u}^{(t + 1)} \leftarrow \bm{u}^{(t)} + \eta_{t + 1}\bm{r}(\bm{u}^{(t)}) + \eta_{t + 1}(\bm{R}(x_{t + 1}) - \bm{r}(\bm{u}^{(t)})),
\end{equation}
in which $R_n(x)  = \mathbbm{1}(E(x) = E_n) - 1/N$, and $r(\bm{u})$ is the mean-field function defined as
\begin{equation}\nonumber
\bm{r}(\bm{u}) = \int_{\mathsf{X}}\bm{R}(x)\pi_{\bm{u}}(x)dx = \frac{\exp(\bm{u}^\star - \bm{u})}{\sum_{n = 1}^N\exp(u_n^\star - u_n)} - \frac{1}{N}.
\end{equation}
The proof of the almost-sure convergence concludes by applying the standard stochastic approximation theory (Theorem 2.2 and Theorem 2.3 in \citep{andrieu2005stability}) after we establish the following two facts. (1) The remainder term $\eta_{t + 1}(\bm{R}(x_{t + 1}) - \bm{r}(\bm{u}^{(t)}))$ vanishes properly as $t\to\infty$. (2) There exists a Lyapunov function $V(\bm{u})$ specified below, 
\begin{equation}
V(\bm{u}) = \frac{1}{N}\sum_{n = 1}^N \exp(u^\star_n - u_n) - 1,
\end{equation}
with respect to the mean-field function $r(\bm{u})$, such that $\langle\nabla V(\bm{u}), \bm{r}(\bm{u})\rangle < 0,\ \forall\ \bm{u}\neq\bm{u}^\star$, and $\langle\nabla V(\bm{u}^\star), \bm{r}(\bm{u}^\star)\rangle = 0$.

The second part of Proposition \ref{prop:convergence-WL} is our main theoretical contribution. There are two essential ingredients in establishing the convergence rate. (i) Strong convexity. The objective function $h(\bm{u})$ is only convex but not strongly convex on $\mathbbm{R}^N$. However, because $\bm{u}^{(t)}$ stays in a compact set $\mathcal{K} \subseteq \Pi$ almost surely (see Proposition \ref{prop:convergence-WL}, part 1(a)), we are able to establish the strong convexity of $h(\bm{u})$ restricted on this compact set $\mathcal{K}$.
\begin{lemma}
\label{lemma:strongly-convex}
Under Assumption \ref{assump:uniform-ergodicity}, there exists a constant $\ell > 0$ such that for any $t \geq 0$, almost surely, it holds
\begin{equation}
\langle\nabla h(\bm{u}^{(t)}), \bm{u}^{(t)} - \bm{u}^\star\rangle \geq \ell ||\bm{u}^{(t)} - \bm{u}^\star||^2.
\end{equation}
\end{lemma}
(ii) Vanishing bias. Because $x_{t + 1}$ is only an approximate sample from the intermediate target distribution $\pi_t(x)$, the indicator $\mathbbm{1}\left(E(x_{t + 1}) = E_n\right)$ is not an unbiased estimator to the negative gradient $\pi_t(E_n)$. The following Lemma \ref{lemma:bias-bound} shows that the bias of the gradient estimator vanishes properly, as fast as the learning rate, when $t\to\infty$.
\begin{lemma}
\label{lemma:bias-bound}
Under Assumption \ref{assump:uniform-ergodicity}, there exists a constant $C > 0$ such that 
\begin{equation}
\mathbbm{E}||\pi_t - P_t(x_t, \cdot)||_{\textnormal{TV}} \leq C\eta_{t + 1}.
\end{equation}
\end{lemma}

The convergence rate of the WL algorithm has been established in different forms in the literature. Zhou and Bhatt \citep{zhou2005understanding} show that the discrete probability distribution $\{\pi_t(E_n)\}_{n = 1}^N$ will be attracted, in terms of the KL-divergence, to the vicinity of the uniform distribution ($\pi_{\infty}(E_n) = 1/N$) as $t\to\infty$. In addition, they show that the standard deviation of $\exp(u_n^\star - u_n^{(t)})$ roughly scales like $\sqrt{\log f_t}$ when the modification factor $f_t$ is close to 1. Although we are looking at the $L^2$ error of $\bm{u}^{(t)}$, which is slightly different from the aforementioned standard deviation, their convergence rate is consistent with our result because $\sqrt{\log f_t}  = \sqrt{\eta_t}$ is in the order of $O(1/\sqrt{t})$ if we scale down the learning rate $\eta_t$ in the order of $O(1/t)$.
It is also worthwhile to mention that a corresponding central limit theorem in the original density space is provided in \citep{fort2015convergence}. 

\section{Accelerating Wang-Landau Algorithm}
\label{sec:acceleration}
The optimization formulation motivates us to further improve the WL algorithm using optimization tools \citep{ruder2016overview}. Our goal in this paper is to accelerate the convergence in the transient phase. The transient phase \citep{darken1992towards} generally refers to the initial stage of running a stochastic gradient descent algorithm. For instance, if we scale down the learning rate according to the flat/minimum histogram criterion, we can refer to the transient phase as the running period from the beginning up to the time when the flat/minimum histogram criterion is first satisfied.

When the transient phase appears noticeable, the acceleration tools can be very effective in practice, and have been widely used in large-scale systems such as deep neural networks \citep{sutskever2013importance}. In this paper, we restrict ourselves on the first-order acceleration methods, and leave other possibilities for future explorations. In particular, we find that both the momentum method and the adaptive learning rate method are effective in accelerating the WL algorithm. Before we go into details, we note that improvement in the asymptotic convergence rate of the stochastic gradient descent algorithm is hard to achieve (or even impossible) \citep{nemirovski2009robust, jain2017accelerating} except for some well-structured objective functions such as finite sums. 

The momentum method exponentially accumulates a momentum vector, denoted as $\bm{m}_t$ in the following, to amplify the persistent gradient across iterations. 
The basic momentum update operates as follows:
\begin{equation}
\label{eq:momentum-update}
\begin{aligned}
& \bm{m}^{(t)} \leftarrow \beta \bm{m}^{(t - 1)} + \eta_{t + 1} \nabla h(\bm{u}^{(t)}), \\
& \bm{u}^{(t + 1)} \leftarrow \bm{u}^{(t)} - \bm{m}^{(t)},
\end{aligned}
\end{equation}
where we initialize the momentum vector to be $\bm{m}^{(0)} = \bm{0}$.
We note that the momentum update essentially adds a fraction $\beta$ of the previously accumulated gradients $\bm{m}^{(t - 1)}$ into the current update vector $\bm{m}^{(t)}$. The weighting factor $\beta$ is a tuning parameter, and is commonly set to be 0.9 or higher. 

In the setting of the WL algorithm, the momentum update in Equation \eqref{eq:momentum-update} becomes
\begin{equation}
\label{eq:momentum-update-WL}
\begin{aligned}
& m_n^{(t)} \leftarrow \beta m_n^{(t - 1)} - \eta_{t + 1} \mathbbm{1}(E(x_{t + 1}) = E_n), \\
& u_n^{(t + 1)} \leftarrow u_n^{(t)} - m_n^{(t)},\ \ \ \ \ \forall n\in[N].
\end{aligned}
\end{equation}
The intuition behind the momentum acceleration for the WL algorithm can be heuristically described as follows. The event $E(x_{t + 1}) = E_n$ suggests that $\pi_t(E_n)$ is likely larger than $1/N$, thus the Markov kernel $P_t$ has a better chance to transit the microscopic configuration $x_t$ into the energy level $E_n$. Therefore, in order to push $\pi_t(E_n)$ towards $1/N$, that is, downweight the probability mass in the energy level $E_n$, we increase $u_n^{(t)}$ by $\eta_{t + 1}$, which corresponds to the density update in Equation \eqref{eq:density-update}. 
In contrast to the WL algorithm, which only increases $u^{(t)}_n$ by $\eta_{t + 1}$ at the current iteration $t$, we keep increasing $u^{(t)}_n$ for a few more iterations by an exponentially decay momentum $m^{(t)}_n$ to achieve a faster convergence. 

The adaptive learning rate method helps standardize the gradient across different coordinates of the parameter $\bm{u}$, so that they scale in a similar magnitude. Otherwise, it can be challenging to find a suitable global learning rate $\eta_t$ over different coordinates.
Popular algorithms along this research direction include AdaGrad \citep{duchi2011adaptive}, AdaDelta \citep{zeiler2012adadelta}, and RMSprop (an unpublished method proposed by Geoffrey Hinton). The RMSprop update operates as follows:
\begin{equation}
\label{eq:RMSprop-update}
\begin{aligned}
& \bm{G}^{(t)} \leftarrow \gamma \bm{G}^{(t - 1)} + (1 - \gamma){\nabla h(\bm{u}^{(t)})}^2, \\
& \bm{u}^{(t + 1)} \leftarrow \bm{u}^{(t)} - {\eta_{t + 1}}[\bm{G}^{(t)}]^{-1/2}\nabla h(\bm{u}^{(t)}),
\end{aligned}
\end{equation}
in which both the square and the square root are taken elementwise. $\bm{G}^{(t)}$ represents the moving average of the squared gradients, so that the current gradient $\nabla h(\bm{u}^{(t)})$, standardized by $[\bm{G}^{(t)}]^{1/2}$, is in a similar magnitude across different coordinates. The weighting factor $\gamma$ is a tuning parameter, which is commonly set to be 0.9 in order to prevent the updates from diminishing too fast. 
In the setting of the WL algorithm, the RMSprop update in Equation \eqref{eq:RMSprop-update} becomes
\begin{equation}
\label{eq:RMSprop-update-WL}
\begin{aligned}
& G^{(t)}_n \leftarrow \gamma G_n^{(t - 1)} + (1 - \gamma)\mathbbm{1}(E(x_{t + 1}) = E_n), \\
& u_n^{(t + 1)} \leftarrow u_n^{(t)} - {\eta_{t + 1}}[G^{(t)}_n]^{-1/2}\mathbbm{1}(E(x_{t + 1}) = E_n).
\end{aligned}
\end{equation}

The combination of the momentum method and the adaptive learning rate method leads to the Adaptive Moment Estimation (Adam) method \citep{kingma2014adam}. The Adam update operates as follows:
\begin{equation}
\label{eq:Adam-update}
\begin{aligned}
& \bm{m}^{(t)} \leftarrow \beta \bm{m}^{(t - 1)} + (1 - \beta) \nabla h(\bm{u}^{(t)}), \\
& \bm{G}^{(t)} \leftarrow \gamma \bm{G}^{(t - 1)} + (1 - \gamma){\nabla h(\bm{u}^{(t)})}^2, \\
& \bm{u}^{(t + 1)} \leftarrow \bm{u}^{(t)} - {\eta_{t + 1}}[\bm{G}^{(t)}]^{-1/2}\bm{m}^{(t)}.
\end{aligned}
\end{equation}
In the setting of the WL algorithm, we note that,  although $\beta$ and $\gamma$ can be potentially two tuning parameters, if we set $\beta = \gamma$ and initialize $\bm{m}^{(0)}$ and $\bm{G}^{(0)}$ to be $\bm{0}$, we have $\bm{G}^{(t)} = - \bm{m}^{(t)}$, since $-\nabla h(\bm{u}^{(t)})$ is approximated by a one-hot vector, which contains only a single ``1" with the remaining elements being 0. This simplification leads to Algorithm \ref{alg:AWL}, which we refer to as the AWL algorithm henceforth.
\begin{algorithm}
\caption{\small{Accelerated Wang-Landau algorithm}}
\label{alg:AWL}
\begin{enumerate}
\item Initialization. $u^{(0)}_n = 0$, $m_n
^{(0)} = 0$ for $n\in[N]$.
\item For $t \geq 1$, iterate between the following steps.
\begin{enumerate}
\item Sample $x_{t + 1}$ from $P_t(x_t, \cdot)$.
\item Update $\bm{m}^{(t)}$ and $\bm{u}^{(t + 1)}$ as follows.
\begin{equation}
\label{eq:Adam-update-WL}
\begin{aligned}
& m^{(t)}_n \leftarrow \beta m^{(t - 1)}_n + (1 - \beta) \mathbbm{1}(E(x_{t + 1}) = E_n), \\
& u^{(t + 1)}_n \leftarrow u^{(t)}_n  + \eta_{t + 1}[m^{(t)}_n]^{1/2}.
\end{aligned}
\end{equation}
\item Normalize $\bm{u}^{(t + 1)}$ to sum to 0.
\item Scale down the learning rate $\eta_t$ properly.
\end{enumerate}
\item Stop when the learning rate $\eta_t$ is smaller than a prescribed threshold.
\end{enumerate}
\end{algorithm}

We remark that for large-scale systems, a naive implementation of Equation \eqref{eq:Adam-update-WL} can be very inefficient, as we have to loop over every coordinate of $\bm{m}^{(t)}$ and $\bm{u}^{(t)}$ in each iteration. A simple solution is to introduce a vector $\bm{s} = (s_1, \cdots, s_N)$, in which $s_n$ records the last time when $m_n$ and $u_n$ are updated. 
With the help of $s_n$, instead of updating $m_n$ and $u_n$ in each iteration, we shall update them only when the energy level $E_n$ is involved in the Monte Carlo simulations.

\section{Illustrations}
\label{sec:illustration}
We compare the AWL algorithm with the original WL algorithm on two benchmark examples: (a) a nearest-neighbour Ising model; (b) a nearest-neighbour ten-state Potts model. Both models are defined on a two-dimensional $L\times L$ square lattice equipped with the periodic boundary condition. 

For the Ising model, the energy $E(x)$ is given by the Hamiltonian:
\begin{equation}
E(x) = -\sum_{<i,j>}J_{ij}x_ix_j - \psi\sum_{j}b_jx_j,
\end{equation}
where $x_i \in\{\pm1\}$. The subscripts $i,j$ denote the lattice sites, and the notation $<i,j>$ implies that the site $i$ and the site $j$ are nearest neighbors. For the ten-state Potts model, the energy $E(x)$ is given by:
\begin{equation}
E(x) = -\sum_{<i,j>}J_{ij}\mathbbm{1}(x_i = x_j) - \psi\sum_{j}b_jx_j,
\end{equation}
where $x_i\in\{1,\cdots, 10\}$. For both models, we assume that $J_{ij}\equiv 1$ and $b_j\equiv0$ (no external magnetic field). If $b_j\equiv0$, the two-dimensional Ising model exhibits a second-order phase transition. Otherwise, in the presence of an external magnetic field, the two-dimensional Ising model exhibits a first-order phase transition. When $b_j\equiv0$, the two-dimensional Potts model exhibits a first-order phase transition when the number of states is larger than 4.

Let $\{H_t(E_n)\}_{n = 1}^N$ be the histogram of all energy levels at iteration $t$. We initialize $H_0(E_n) = 0$ for $n\in[N]$. At each iteration $t$, the AWL algorithm and the WL algorithm update $\bm{u}^{(t)}$ according to Algorithm \ref{alg:AWL} and Algorithm \ref{alg:WL}, respectively. In addition, we update the energy histogram as $H_t(E_n) = H_{t - 1}(E_n) + \mathbbm{1}(E(x_{t + 1}) = E_n)$. 

The adaptation of the learning rate $\eta_t$ follows \citep{belardinelli2007fast}, which is detailed in the following.
\begin{enumerate}
\item After every 1,000 MC sweeps, we check $\{H_t(E_n)\}$. If $\min_n H_t(E_n) > 0$, we set $\eta_{t + 1} = \eta_t/2$, and reset $H_t(E_n) = 0$ for each energy level $E_n$. Otherwise if $\min_n H_t(E_n) = 0$, we keep $\eta_{t + 1} = \eta_t$.
\item If $\eta_{t + 1}\leq N/t$, then $\eta_t = N/t$ for all the subsequent iterations. $H_t(E_n)$ is discarded and the above step is not executed any more.
\end{enumerate}
We note that each MC sweep contains $L^2$ iterations, in which each iteration refers to a single round of parameter update. That is, step 2(a)--2(c) in Algorithm \ref{alg:WL} and Algorithm \ref{alg:AWL}.
The energy histogram $\{H_t(E_n)\}$ essentially represents the number of visits to each energy level up to iteration $t$, since the last update of the learning rate. 

We implement one step of the Metropolis algorithm to estimate the gradient, i.e., step 2(a) in Algorithm \ref{alg:WL} and Algorithm \ref{alg:AWL}. The proposal schemes for the Ising model and the Potts model are described as follows. Given the current configuration $x_t$, we randomly pick up a site and change its value. For the Ising model, we filp its sign. For the ten-state Potts model, we set it to be a number uniformly sampled from $\{1, \cdots, 10\}$.

To illustrate the efficiency of the AWL algorithm, we investigate the following four perspectives. (i) The scaling of the first equilibration time, in terms of the number of MC sweeps, with respect to the dimension $L$. The first equilibration time, which corresponds to the transient phase as we discussed in Section \ref{sec:acceleration}, is defined to be $\min\{t: \min_n H_t(E_n)\} > 0$. That is, the first time when the energy histogram becomes nonzero everywhere. According to the adaptation rule of the learning rate $\eta_t$, the equilibration time is also the first time we decrease the learning rate. (ii) The scaling of the first equilibration time, in terms of the CPU time, with respect to the dimension $L$. Because the AWL algorithm requires additional computations in updating the momentum vector, the comparison between the two algorithms on the actual CPU time is necessary to see whether the implementation of the acceleration method is indeed worthwhile. (iii) The dynamics of the estimation error $\epsilon(t)$ defined as below following \citep{belardinelli2007fast} for $L = 80$,
\begin{equation}
\epsilon(t) = \frac{1}{N - 1}\sum_{n = 1}^N \left|1 - \frac{\log(g_t(E_n))}{\log(g(E_n))}\right|.
\end{equation}
For the Ising model, the exact density of states $g(E_n)$ is available, and can be calculated using a publicly available Mathematica program \citep{beale1996exact}. For the Potts model, no exact solution of $g(E_n)$ is available, thus we pre-run a $1/t$ WL simulation for $5\times10^{7}$ MC sweeps, in which the final learning rate is $2\times10^{-8}$. We then treat the density estimates as an approximation to the exact density of states. 
(iv) The accuracy in the task of estimating the specific heat for the Ising model with $L = 80$.

We compare the AWL algorithm and the WL algorithm with different initializations of the learning rate, $\eta_0 = 0.05$, $0.10$ and $1.00$. We test out the two algorithms for different sizes of the two-dimensional square lattice, $L = 50, 60, 70, 80, 90, 100$. 
The computations in this paper were run on the FASRC Cannon cluster supported by the FAS Division of Science Research Computing Group at Harvard University.

\begin{figure*}
\centering
\includegraphics[width=1.0\columnwidth]{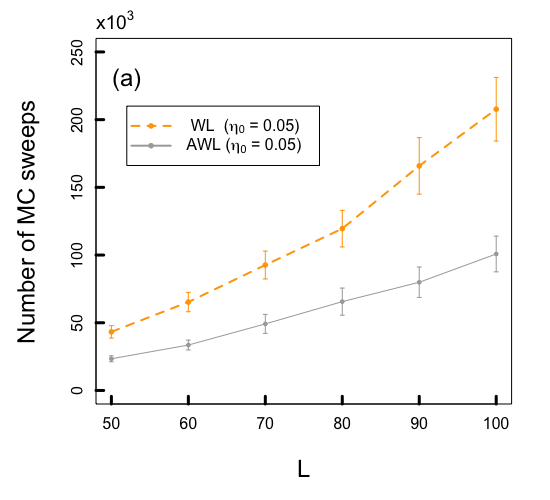}
\includegraphics[width=1.0\columnwidth]{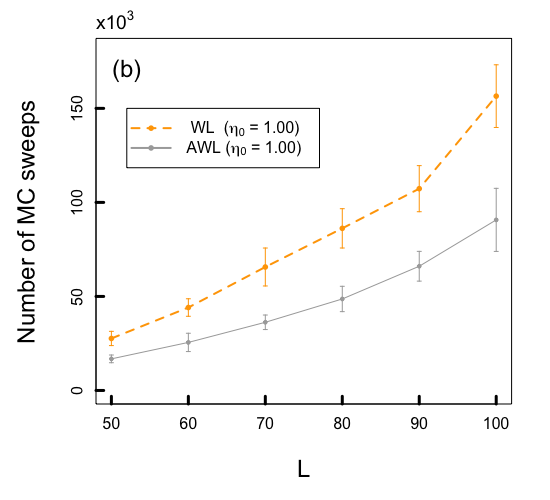}
\includegraphics[width=1.0\columnwidth]{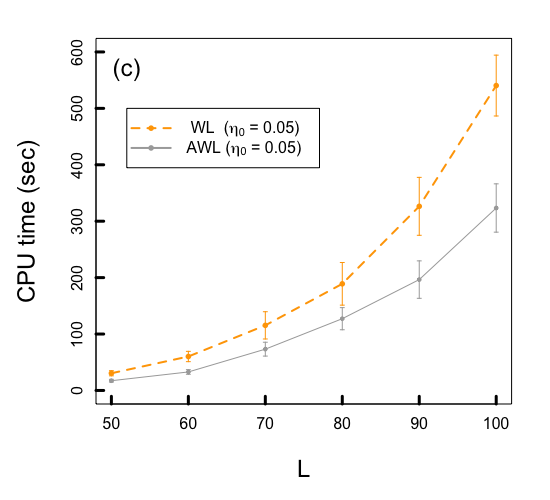}
\includegraphics[width=1.0\columnwidth]{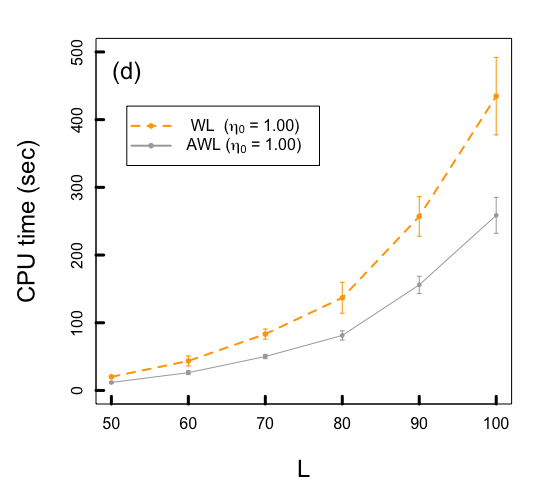}
\caption{The computational overheads, in terms of the number of MC sweeps and the CPU time, that the AWL algorithm and the WL algorithm takes to reach the first equilibration on the Ising model. Two initializations of the learning rate are tested out, including $\eta_0 = 0.05$ and $\eta_0 = 1.00$. The reported results are based on 50 independent runs of both algorithms. The dot represents the empirical mean and the error bar represents the empirical standard deviation.}
\label{fig:Ising-computation}
\end{figure*}

\begin{figure*}
\centering
\includegraphics[width=1.0\columnwidth]{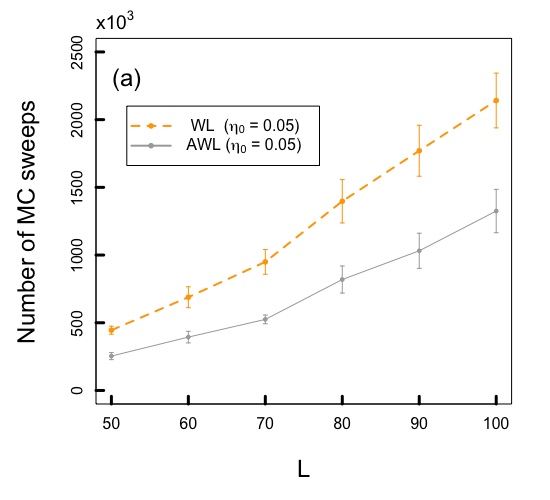}
\includegraphics[width=1.0\columnwidth]{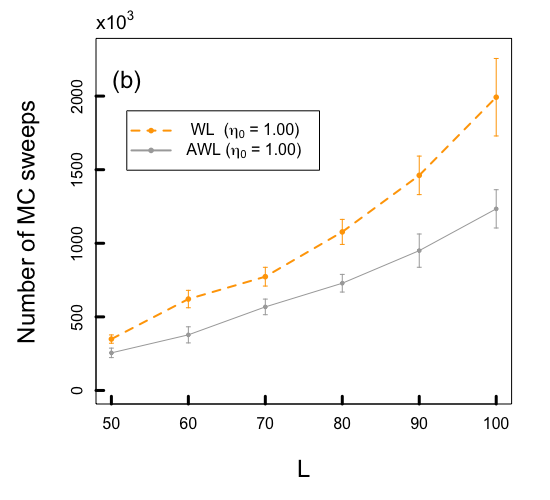}
\includegraphics[width=1.0\columnwidth]{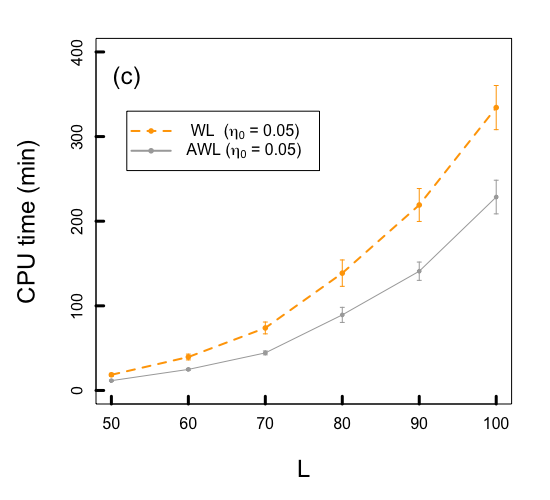}
\includegraphics[width=1.0\columnwidth]{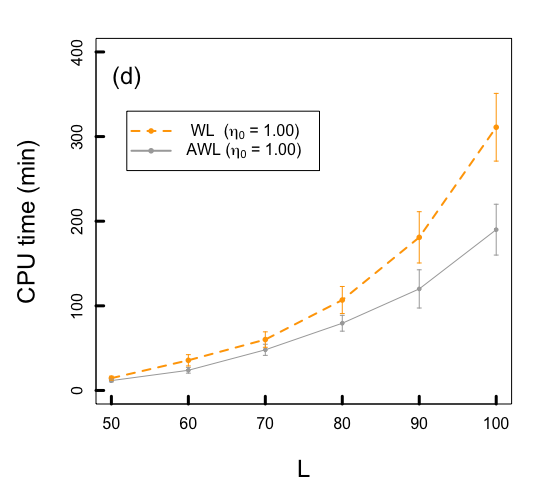}
\caption{The computational overheads, in terms of the number of MC sweeps and the CPU time, that the AWL algorithm and the WL algorithm takes to reach the first equilibration on the Potts model. Two initializations of the learning rate are tested out, including $\eta_0 = 0.05$ and $\eta_0 = 1.00$. The reported results are based on 50 independent runs of both algorithms. The dot represents the empirical mean and the error bar represents the empirical standard deviation.}
\label{fig:Potts-computation}
\end{figure*}

Figure \ref{fig:Ising-computation} summarizes the computational overheads of the two algorithms for the Ising model. The reported results are based on 50 independent runs of both algorithms, in which the dot represents the empirical mean and the error bar represents the empirical standard deviation. We see that the AWL algorithm takes significantly fewer MC sweeps as well as less CPU time to reach the first equilibration among all settings with different lattice sizes and different initializations of the learning rate. 
Figure \ref{fig:Potts-computation} summarizes the computational overheads of the two algorithms on the Potts model. Similar to the case of Ising model, the AWL algorithm is more efficient than the WL algorithm in terms of the first equilibration time measured by the number of MC sweeps and the CPU time. 

\begin{figure*}
\centering
\includegraphics[width=1.0\columnwidth]{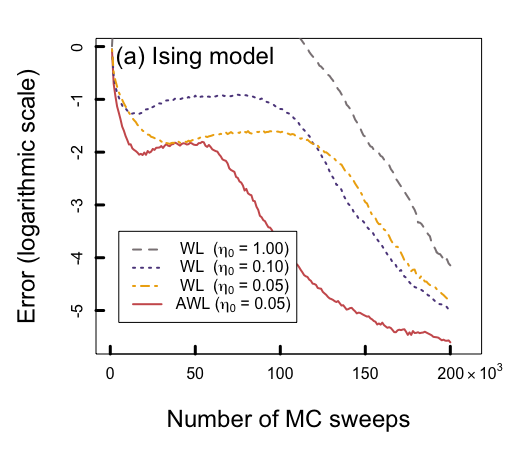}
\includegraphics[width=1.0\columnwidth]{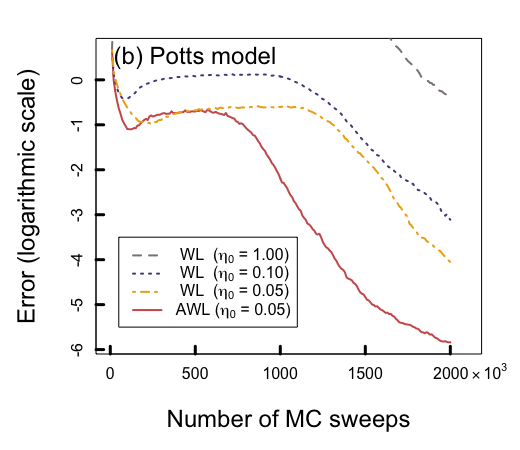}
\caption{The dynamics of the estimation error $\epsilon(t)$ (in the logarithmic scale), averaging over 50 independent runs, of the AWL algorithm and the WL algorithm. Panel (a) shows the result for the Ising model, and panel (b) shows the result for the Potts model. $\eta_0$ denotes the initialization of the learning rate. }
\label{fig:error}
\end{figure*}

Figure \ref{fig:error} shows the empirical dynamics of $\epsilon(t)$, averaged over 50 independent runs of both algorithms. The first $100\times10^3$ MC sweeps for the Ising model and the first $1500\times 10^3$ MC sweeps for the Potts model are representative for the transient phase. We see that in the transient phase, the convergence speed of the AWL algorithm, in terms of the number of MC sweeps,  is significantly faster than the convergence speed of the WL algorithm with different initializations of the learning rate. 

For the Ising model with $L = 80$, Table \ref{tab:specific-heat} compares the accuracy of the two algorithms in the calculation of the specific heat defined as:
\begin{equation}
\label{eq:specific_heat}
C(T) = \frac{\langle E^2\rangle_T - \langle E\rangle_T^2}{T^2},
\end{equation}
in which $T$ denotes the temperature. We test out temperatures ranging from 0.4 to 8 incremented by 0.1. The internal energy $\langle E\rangle_T$ is defined as 
\begin{equation}
\begin{aligned}
\langle E\rangle_T & = \frac{\sum_{n}E_ng(E_n)\exp(-E_n/T)}{\sum_{n}g(E_n)\exp(-E_n/T)}.
\end{aligned}
\end{equation}
The fluctuation expression $\langle E^2\rangle_T$ is defined similarly. We note that the theoretical value of the specific heat at a given temperature $T$ can be evaluated exactly when the exact density of states is available, which is the case for the two-dimensional Ising model. We independently run each algorithm 50 times to obtain 50 independent estimates of the specific heat at each temperature. The relative error at each temperature is calculated based on the mean of the 50 independent estimates. Table \ref{tab:specific-heat} summarizes the quantiles of the relative errors for $T\in[0.4, 8]$, by running each algorithm for $100\times10^3$, $150\times10^3$, and $200\times10^3$ MC sweeps, respectively. Compared to the WL algorithm, the AWL algorithm yields significantly more accurate estimates of the specific heat especially in the transient phase.
 
\begin{table*}
\begin{center}
\begin{tabularx}{0.8\textwidth}{c *{9}{Y}}
\tabularnewline \toprule[0.8pt]
 & \multicolumn{3}{c}{$100\times10^3$ MC sweeps}  
 & \multicolumn{3}{c}{$150\times10^3$ MC sweeps}
  & \multicolumn{3}{c}{$200\times10^3$ MC sweeps}
\tabularnewline  \cmidrule[0.2pt](lr){2-4} \cmidrule[0.2pt](l){5-7}\cmidrule[0.2pt](l){8-10} 
Quantiles & 25$\%$ & $50\%$ & $75\%$ & 25$\%$ & $50\%$ & $75\%$ & 25$\%$ & $50\%$ & $75\%$ 
\tabularnewline \midrule  [0.5pt]%
AWL ($\eta_0 = 0.05$) & $2.9\%$ & $6.3\%$ & $17.7\%$ & $0.9\%$ & $2.0\%$ & $4.6\%$ & $0.5\%$ & $1.2\%$ & $2.9\%$\\
\hspace{0.1cm} WL ($\eta_0 = 0.05$) & $10.5\%$ & $18.9\%$ & $41.4\%$ & $4.6\%$ & $9.1\%$ & $17.7\%$ & $1.1\%$ & $2.0\%$ & $4.4\%$\\
\hspace{0.1cm} WL ($\eta_0 = 0.10$) & $12.2\%$ & $24.0\%$ & $44.0\%$ &$2.4\%$ & $4.6\%$ & $10.9\%$ & $0.7\%$ & $2.4\%$ & $5.1\%$\\
\hspace{0.1cm} WL ($\eta_0 = 1.00$) & $47.1\%$ & $57.6\%$ & $74.4\%$ & $8.0\%$ & $16.0\%$ & $27.5\%$ & $2.8\%$ & $4.6\%$ & $8.4\%$
\tabularnewline\bottomrule [0.8pt]%
\end{tabularx}
\end{center}
\caption{The relative errors of the AWL algorithm and the WL algorithm in the calculation of the specific heat for the Ising model with $L = 80$. The relative errors are calculated based on the mean of 50 independent estimates produced by each algorithm. The quantiles of the relative errors are over the temperature interval $T\in[0.4, 8]$. $\eta_0$ denotes the initialization of the learning rate.}\label{tab:specific-heat}
\end{table*}

More details of this numerical study can be found in the Supplemental Material. First, within the first $2\times10^5$ MC sweeps and $2\times10^6$ MC sweeps for the Ising model and the Potts model, respectively, we report the number of equilibrations that the AWL algorithm and the WL algorithm have reached (equivalently, the number of changes of the learning rate $\eta_t$), for different lattice sizes $L$ and different initializations of the learning rate $\eta_0$. We also report the corresponding first 8 equilibration time in terms of the number of MC sweeps. Second, for the Ising model with $L = 80$, we provide a graphical comparison of the estimated specific heat obtained by the AWL algorithm and the WL algorithm, over the temperature region $T\in[0.4, 8]$.

\section{Conclusion}
\label{sec:conclusion}
To summarize, in this paper we present a new interpretation of the WL algorithm from the optimization perspective. We show that the WL algorithm is essentially a stochastic (projected) gradient descent algorithm minimizing a smooth and convex function, in which MCMC steps are used to estimate the unknown gradient. 
The optimization formulation intuitively explains that because of using more accurate gradient estimates, some notable modifications of the algorithm, such as utilizing multiple random walkers, can improve the WL algorithm. In addition, using the (strong) convexity of the objective function, we provide a new approach to establish the convergence rate of the WL algorithm, which is more explicit compared to the existing results \citep{fort2015convergence, zhou2005understanding}. We expect that our contributions are useful for further theoretical investigations of the WL algorithm.

The optimization interpretation also opens a new way to improve the efficiency of the WL algorithm. There are rich tools in the optimization literature to accelerate the stochastic gradient descent algorithm, including but not restricted to the methods we mentioned in Section \ref{sec:acceleration}. Different methods can be favorable for different applications. In the presence of noisy gradients, it usually requires some careful tuning to successfully apply the acceleration tools. We demonstrate  one possible acceleration approach, using the momentum method and the adaptive learning rate strategy, on a two-dimensional Ising model and a two-dimensional ten-state Potts model.

\textit{Acknowledgments}. This work was partially supported by NSF DMS-1613035
and DMS-1712714.

\bibliographystyle{unsrt}
\bibliography{AWL.bib}
\end{document}